\documentstyle{mn}
\font\japit = cmti10 at 10truept
\input{epsf.sty}

\title
     [Optimal Power Spectra II]
{\vglue-3.0truecm
\centerline{\japit Submitted to Monthly Notices}
\vglue 2.5truecm
\noindent
Towards Optimal Measurement of Power Spectra II: \\
A Basis of Positive, Compact, Statistically Orthogonal Kernels
\author
     [A. J. S. Hamilton]
     {A. J. S. Hamilton \\
	JILA and Dept.\ of Astrophysical, Planetary and Atmospheric Sciences,
	Box 440, University of Colorado, Boulder CO 80309, USA\\
	email: ajsh@dark.colorado.edu}}

\newcommand{\rmn}{\rm}
\newcommand{\bmi}{\bmath}
\newcommand{\be}{\begin{equation}}
\newcommand{\ee}{\end{equation}}
\newcommand{\ba}{\begin{eqnarray}}
\newcommand{\ea}{\end{eqnarray}}
\newcommand{\nn}{\nonumber \\}
\newcommand{\dd}{{\rmn d}}	% MNRAS
\newcommand{\e}{{\rmn e}}	% MNRAS
\newcommand{\im}{{\rmn i}}	% MNRAS
\newcommand{\ddd}{\dd^3\!}

\newcommand{\simpropto}{\!\!\begin{array}{c} {\propto} \\
                  [-1.7ex] \sim \end{array}\!\!}

\newcommand{\k}{{\bmi k}}

\newcommand{\r}{{\bmi r}}

\newcommand{\Mpc}{{\rmn Mpc}}

\newcommand{\ff}{\beta}
\newcommand{\tildeR}{\tilde R}
\newcommand{\PI}{\pi}

\begin{document}

\maketitle

\begin{abstract}
This is the second of two papers which address
the problem of measuring the unredshifted power spectrum of fluctuations
from a galaxy survey in optimal fashion.
A key quantity is the Fisher matrix,
which is the inverse of the covariance matrix of minimum variance
estimators of the power spectrum of the survey.
It is shown that bases of kernels which give rise to complete
sets of statistically orthogonal windowed power spectra
are obtained in general from the eigenfunctions of the Fisher matrix
scaled by some arbitrary positive definite scaling matrix.
Among the many possible bases of kernels,
there is a basis, obtained by applying an infinitely steep scaling function,
which leads to kernels which are positive and compact in Fourier space.
This basis of kernels, along with the associated minimum variance pair
weighting derived in the previous paper, would appear to offer a solution
to the problem of how to measure the unredshifted power spectrum optimally.
Illustrative kernels are presented for the case of the PSCz survey.
\end{abstract}


%\clearpage

\section{Introduction}
\label{intro}

This is the second of a pair of papers
(Hamilton 1997, hereafter Paper~1)
concerned with the problem of measuring the unredshifted power spectrum
of fluctuations in a galaxy survey in optimal fashion.

As discussed by Tegmark, Taylor \& Heavens (1997),
a fundamental quantity in the estimation of any set of parameters
from a data set is the Fisher information matrix,
which is the matrix of second derivatives of minus the log-likelihood function
with respect to the parameters,
and whose inverse gives an estimate of the covariance matrix of the parameters.
It was demonstrated in Paper~1 that if the parameter set of interest
is the unredshifted power spectrum of fluctuations in a galaxy survey,
and under the standard assumption that the main source of uncertainty
in the power spectrum is cosmic variance plus Poisson sampling noise,
then the Fisher matrix can be evaluated rather directly,
albeit in an approximation.
With the Fisher matrix in hand,
it becomes possible to construct complete sets of kernels
which yield statistically orthogonal sets of windowed power spectra.
Such bases of kernels hold the potential of providing an optimal way
to measure the power spectrum,
in the sense that they offer
statistical orthogonality
(which implies statistical independence in the case of Gaussian fluctuations),
and completeness
(hence maximum resolution,
since the power spectrum cannot be decomposed further into statistically
orthogonal subunits).
The purpose of the present paper is to explore this idea further,
and in particular to show some actual examples.
I choose to use the geometry and selection function of PSCz survey
(W.\ Saunders et al., in preparation)
for illustration,
and I assume Gaussian fluctuations for simplicity.

As shown in \S\ref{orthog},
there are in fact many distinct bases of kernels which give rise to
statistically orthogonal sets of windowed power spectra.
After giving details of the computation of the Fisher matrix in \S\ref{compute},
I show two trial bases of kernels in \S\ref{trial}.
It is found there that, without special attention,
kernels are liable to be broad and wiggly in Fourier space,
which are not desirable properties.
The difficulty is resolved in \S\ref{compact},
where it is shown how to construct a basis of kernels which are
compact and everywhere positive in Fourier space.
Section~\ref{compact} leaves unanswered a number of basic questions
as to the generality and uniqueness of the proposed construction.
These questions are posed \S\ref{questions}.
The results are summarized in \S\ref{summary}.

\section{Statistically Orthogonal Sets of Power Spectra}
\label{orthog}

Let $\xi_\alpha$ denote the power spectrum $\xi(k)$
windowed through a set of kernels $\psi_\alpha(k)$:
\be
\label{xia}
  \xi_\alpha \equiv
    \int_0^\infty \psi_\alpha(k) \xi(k) \, 4\PI k^2 \dd k / (2\PI)^3
  \ .
\ee
The power spectrum $\xi(k)$ is real,
and the kernels $\psi_\alpha(k)$ can also be taken to be real without
loss of generality.
Let $\hat\xi_\alpha$ denote the minimum variance estimate of
the windowed power spectrum $\xi_\alpha$
(as in Paper~1, the hat on $\hat\xi$ is used to distinguish
the estimator from the true power spectrum $\xi$)
\ba
  \tilde\xi_\alpha
    \!\!\! &\approx& \!\!\!
    \hat\xi_\alpha
  \ ,
  \nn
\label{xihata}
  \hat\xi_\alpha
    \!\!\! &\equiv& \!\!\!
    \int W(\r_i,\r_j) \delta(\r_i) \delta(\r_j) \, \ddd r_i \ddd r_j
\ea
where $W(\r_i,\r_j)$ is the minimum variance pair window
derived in Paper~1, equation~(39),
for the power spectrum windowed through the kernel $G(k) = \psi_\alpha(k)$.
The integration in equation~(\ref{xihata})
is over all pairs of observed overdensities
$\delta(\r_i) \delta(\r_j)$ at positions $\r_i$ and $\r_j$ in the survey.
The estimators $\hat\xi_\alpha$
of the windowed power spectra $\xi_\alpha$
are said to be statistically orthogonal
if the expectation value of the covariance between different elements vanishes,
\be
\label{xiaxib}
  \langle \Delta\hat\xi_\alpha \Delta\hat\xi_\beta \rangle =
    0
  \quad (\alpha \neq \beta)
  \ .
\ee
Statistical orthogonality does not imply statistical independence in general,
but it does imply statistical independence for the particular case of
Gaussian fluctuations,
for which by definition all the higher order irreducible moments vanish.

There are not one but many distinct bases,
i.e.\ complete linearly independent sets, of kernels $\psi_\alpha$
which define statistically orthogonal sets of power spectra $\hat\xi_\alpha$.
To see this,
consider first that the eigenfunctions of the expected covariance matrix
$\langle \Delta\hat\xi(k) \Delta\hat\xi(k') \rangle$
of estimators $\hat\xi(k)$ of the power spectrum
define a certain complete orthonormal set of kernels $\psi_\alpha(k)$,
for which by construction the windowed power spectra $\hat\xi_\alpha$
are statistically orthogonal.
Now rescale each kernel $\psi_\alpha(k)$ to unit variance,
by dividing it by the square root of the variance,
$\psi_\alpha \rightarrow
\psi_\alpha/\langle \Delta\hat\xi_\alpha^2 \rangle^{1/2}$.
The covariance matrix of the power spectrum
windowed through the resulting basis of scaled kernels is the unit matrix.
Any orthogonal rotation amongst these scaled kernels yields another basis
of kernels,
which defines another distinct statistically orthogonal set
of windowed power spectra.

If $\psi_\alpha(k)$ is a set of kernels which defines a statistically
orthogonal set of estimators $\hat\xi_\alpha$ of windowed power spectra,
then evidently $c_\alpha \psi_\alpha(k)$,
where $c_\alpha$ is any set of finite nonzero normalization constants,
defines a renormalized version of essentially the same
statistically orthogonal set,
with $\hat\xi_\alpha \rightarrow c_\alpha \hat\xi_\alpha$
and
$\langle \Delta\hat\xi_\alpha^2 \rangle \rightarrow
c_\alpha^2 \langle \Delta\hat\xi_\alpha^2 \rangle$.
In the development below it is informative
to carry along a set of normalization constants $c_\alpha$.
The normalization constants can be chosen arbitrarily,
so that one is free to normalize the kernels $\psi_\alpha(k)$
in whatever manner seems convenient.
In the present paper I choose to plot the kernels
quadratically normalized, so that their squares have unit volume in $k$-space,
\be
\label{norm}
  \int_0^\infty \psi_\alpha(k)^2 \, 4\PI k^2 \dd k / (2\PI)^3 = 1
  \ .
\ee
In practical applications,
notably when using the basis of positive kernels
presented in \S\ref{compact},
one might prefer to normalize the kernels linearly
to unit volume in $k$-space,
$\int_0^\infty \psi_\alpha(k) 4\PI k^2 \dd k / (2\PI)^3 = 1$.
The difficulty with the linear normalization
-- more a matter of plotting rather than anything fundamental --
is that for kernels
$\psi_\alpha(k)$ which are not everywhere positive in $k$-space,
such as those considered in \S\ref{trial},
a linear rather than quadratic normalization tends to make noisy
kernels wiggle more fiercely, obscuring the better-behaved kernels.

In a moment I will prove the following theorem.
For any complete set of kernels $\psi_\alpha(k)$,
along with any set of normalization constants $c_\alpha$,
yielding a statistically orthogonal set of windowed power spectra
$\hat\xi_\alpha$,
there exists a strictly positive definite
-- i.e., all the eigenvalues are strictly positive --
real symmetric scaling matrix $\gamma(k,k')$
such that the scaled kernels $\phi_\alpha(k)$ defined by
\be
\label{phi}
  \phi_\alpha(k) \equiv
    \int_0^\infty \gamma(k,k') c_\alpha \psi_\alpha(k')
    \, 4\PI k'^2 \dd k' / (2\PI)^3
\ee
form a complete orthonormal basis
which diagonalizes the scaled covariance matrix
$\gamma^{-1} \! \langle \Delta\hat\xi \Delta\hat\xi \rangle \gamma^{-1}$.

The converse of this theorem is evident.
If $\gamma$ is any strictly positive definite real symmetric matrix,
then the scaled covariance matrix
$\gamma^{-1} \! \langle \Delta\hat\xi \Delta\hat\xi \rangle \gamma^{-1}$
defines a complete orthonormal set of eigenfunctions $\phi_\alpha$,
with eigenvalues $\lambda_\alpha$:
\be
  \gamma^{-1} \!
    \langle \Delta\hat\xi \Delta\hat\xi \rangle
    \gamma^{-1} \phi_\alpha
    = \lambda_\alpha \phi_\alpha
  \ .
\ee
The kernels $\psi_\alpha$ defined by
$\psi_\alpha = c_\alpha^{-1} \gamma^{-1} \! \phi_\alpha$,
the inverse of equation~(\ref{phi}),
then yield estimators $\hat\xi_\alpha$ of
windowed power spectra, equations~({\ref{xia}) and (\ref{xihata}),
which are statistically orthogonal because
\[
  \langle \Delta\hat\xi_\alpha \Delta\hat\xi_\beta \rangle
  \equiv
  \psi_\alpha \langle \Delta\hat\xi \Delta\hat\xi \rangle \psi_\beta
  =
  c_\alpha^{-1} \phi_\alpha \gamma^{-1} \!
    \langle \Delta\hat\xi \Delta\hat\xi \rangle
    \gamma^{-1} \! \phi_\beta c_\beta^{-1}
\]
\be
\label{DxiDxi}
  \quad\quad\quad\quad\ \,
  =
    c_\alpha^{-2} \lambda_\alpha \delta_{D \alpha \beta}
\ee
is a diagonal matrix (no summation over $\alpha$ or $\beta$ implied).
Here $\delta_{D \alpha \beta}$ denotes the unit matrix.
In particular,
the variances
$\langle \Delta\hat\xi_\alpha^2 \rangle$
of the windowed power spectra $\hat\xi_\alpha$
are related to the eigenvalues $\lambda_\alpha$ of
the scaled covariance matrix
$\gamma^{-1} \! \langle \Delta\hat\xi \Delta\hat\xi \rangle \gamma^{-1}$
by
\be
\label{eigval}
  \langle \Delta\hat\xi_\alpha^2 \rangle = {\lambda_\alpha \over c_\alpha^2}
  \ .
\ee

As seen in Paper~1,
what one actually calculates for a survey is not the expected covariance matrix
$\langle \Delta\hat\xi \Delta\hat\xi \rangle$
of estimates of the power spectrum,
but rather its inverse the Fisher matrix
$T = \langle \Delta\hat\xi \Delta\hat\xi \rangle^{-1}$.
The inverse of the scaled covariance matrix
$\gamma^{-1} \! \langle \Delta\hat\xi \Delta\hat\xi \rangle \gamma^{-1}$
is the scaled Fisher matrix
$\gamma T \gamma$.
Thus the eigenfunctions $\phi_\alpha$
of the scaled covariance matrix can in practice be constructed
as the orthonormal eigenfunctions of the scaled Fisher matrix,
\be
\label{eigeq}
  \gamma T \gamma \phi_\alpha = \mu_\alpha \phi_\alpha
  \ ,
\ee
with eigenvalues $\mu_\alpha = 1/\lambda_\alpha$
which are the reciprocals of the eigenvalues of the scaled covariance matrix.
The kernels $\psi_\alpha$ follow as before from
$\psi_\alpha = c_\alpha^{-1} \gamma^{-1} \! \phi_\alpha$,
the inverse of equation~(\ref{phi}).
The variances
$\langle \Delta\hat\xi_\alpha^2 \rangle$
of the windowed power spectra $\hat\xi_\alpha$
are related to the eigenvalues $\mu_\alpha$ of the scaled Fisher matrix by,
equation~(\ref{eigval}),
\be
\label{eigvalT}
  \langle \Delta\hat\xi_\alpha^2 \rangle =
    {1 \over c_\alpha^2 \mu_\alpha}
  \ .
\ee

Thus the general procedure for constructing complete bases of
kernels yielding statistically orthogonal sets of power spectra
is as follows:
Choose any strictly positive definite symmetric scaling matrix $\gamma$;
find the eigenfunctions $\phi_\alpha$ of the scaled Fisher matrix
$\gamma T \gamma$;
and construct the kernels $\psi_\alpha \propto \gamma^{-1} \! \phi_\alpha$,
normalizing to taste.

The next few paragraphs prove the previously stated theorem,
which asserts that a scaling matrix $\gamma$ exists for any complete set of
kernels yielding a statistically orthogonal set of power spectra,
and which therefore guarantees that the procedure of the previous paragraph
is in fact general.
Suppose then that $\psi_\alpha(k)$ comprise a complete set of kernels
which give rise to a statistically orthogonal set $\hat\xi_\alpha$
of windowed power spectra,
equations~(\ref{xia}), (\ref{xihata}), and (\ref{xiaxib}).
The kernels $\psi_\alpha(k)$
themselves need not be orthogonal,
but they must be linearly independent.
For if one of the kernels depended linearly on the others,
say $\psi_\alpha = \sum_{\beta \neq \alpha} a_\beta \psi_\beta$
for some $a_\beta$,
then the corresponding windowed power spectrum would likewise depend linearly
on the others,
$\hat\xi_\alpha = \sum_{\beta \neq \alpha} a_\beta \hat\xi_\beta$,
so that
$\langle \Delta\hat\xi_\alpha \Delta\hat\xi_\beta \rangle
= a_\beta \langle \Delta\hat\xi_\beta^2 \rangle$
would be nonzero for some $\beta$,
contradicting the statistical orthogonality of the $\hat\xi_\alpha$.
Linear independence, coupled with the assumption of completeness,
means that the kernels $\psi_\alpha(k)$ form a basis.

Define the real symmetric matrix $A(k,k')$ by
\be
\label{A}
  A(k,k') \equiv
    \sum_\alpha c_\alpha^2 \psi_\alpha(k) \psi_\alpha(k')
  \ .
\ee
Being real and symmetric,
the matrix $A$ is diagonalized by some orthogonal transformation $O$,
so that $O^T \!\! A O$ is diagonal.
Rotating equation~(\ref{A}) by the orthogonal transformation $O$,
one concludes that the matrix $A$ must be positive definite --
that is, all its eigenvalues are positive or zero --
since each eigenvalue is the inner product of a vector $O c \psi$ with itself.
The assumption that the kernels $\psi_\alpha$ form a complete set
implies further that the eigenvalues of $A$ are all strictly positive.
For equations~(\ref{xia}) and (\ref{A}) together imply
\be
\label{Axi}
  \int_0^\infty A(k,k') \xi(k') \, 4\PI k'^2 \dd k' / (2\PI)^3 =
    \sum_\alpha c_\alpha^2 \psi_\alpha(k) \xi_\alpha
  \ ,
\ee
and if there existed a nontrivial eigenfunction $\xi(k)$ of $A$ with zero
eigenvalue,
then the left hand side of equation~(\ref{Axi}) would vanish
for this eigenfunction.
The consequent vanishing of the right hand side of equation~(\ref{Axi}) would
then imply either that the kernels $\psi_\alpha(k)$ are linearly dependent,
which is false as shown above,
or else that $\xi_\alpha = 0$ for all $\alpha$
for the eigenfunction $\xi(k)$.
But if, as postulated, the kernels $\psi_\alpha(k)$ form a complete set,
then $\xi(k) = \sum_\alpha x_\alpha \psi_\alpha(k)$
for some set of coefficients $x_\alpha$,
so that
$\int \xi(k)^2 4\PI k^2 \dd k / (2\PI)^3 =
\sum_\alpha x_\alpha
\int \psi_\alpha(k) \xi(k) 4\PI k^2 \dd k / (2\PI)^3 =
\sum_\alpha x_\alpha \xi_\alpha = 0$,
whose only possible solution is $\xi(k) = 0$.
Thus the matrix $A$ can have no nontrivial eigenfunctions with zero eigenvalue,
and hence $A$ must be strictly positive definite, as claimed.

It follows that the matrix $A$ of equation~(\ref{A}) can be written in the form
$A = O \Lambda^2 O^T = (O \Lambda O^T)^2 = \gamma^{-2}$
where $\Lambda$ is a diagonal matrix,
and $\gamma \equiv O \Lambda^{-1} O^T$.
Without loss of generality,
all the square roots in $\Lambda = (\Lambda^2)^{1/2}$
can be taken to be positive.
Thus the matrix $\gamma$, which I call the scaling matrix,
is also a strictly positive definite real symmetric matrix.
It then follows from equation~(\ref{A}) that the scaled kernels
$\phi_\alpha(k)$ defined by equation~(\ref{phi})
form a complete orthonormal set
\be
\label{phiphi}
  \sum_\alpha \phi_\alpha(k) \phi_\alpha(k')
    = (2\PI)^3 \delta_D(k-k')
\ee
the Dirac delta-function $(2\PI)^3 \delta_D(k-k')$
being the unit matrix in the Fourier representation.
In accordance with the convention of Paper~1,
the Dirac delta-function in equation~(\ref{phiphi})
is normalized on a 3-dimensional scale, so that
$\int \delta_D(k-k') 4\PI k'^2 \dd k' = 1$.
Equation~(\ref{phiphi})
shows that $\phi_\alpha(k)$ is in effect
an orthogonal matrix $\phi_{k\alpha}$,
whose inverse is its transpose.
Taking the factors in opposite order yields the equivalent orthonormality
condition
\be
\label{phiaphib}
  \int_0^\infty \phi_\alpha(k) \phi_\beta(k) \, 4\PI k^2 \dd k / (2\PI)^3
    = \delta_{D \alpha \beta}
\ee
where $\delta_{D \alpha \beta}$ is the unit matrix
in the $\phi$-representation.

The desired scaling matrix $\gamma$ has thus been constructed,
as asserted by the theorem,
with the property that the scaled kernels $\phi_\alpha$
defined by equation~(\ref{phi}) form a complete orthonormal set.
The final assertion of the theorem,
that the scaled covariance matrix
$\gamma^{-1} \! \langle \Delta\hat\xi \Delta\hat\xi \rangle \gamma^{-1}$
is diagonal with respect to the scaled kernels $\phi_\alpha$,
follows directly, equation~(\ref{DxiDxi}), from the hypothesis that
the covariance matrix $\langle \Delta\hat\xi \Delta\hat\xi \rangle$
is diagonal with respect to the kernels $\psi_\alpha$.
This completes the proof of the theorem.

From the definition~(\ref{xia}) of $\xi_\alpha$,
and the orthonormality condition~(\ref{phiphi})
on the scaled kernels $\phi_\alpha(k)$ defined by equation~(\ref{phi}),
it follows that the power spectrum $\xi(k)$ can be expanded
in terms of the kernels $\psi_\alpha(k)$ as
\be
\label{xik}
  \xi(k) = \sum_\alpha \xi_\alpha
    c_\alpha^2
    \int_0^\infty \gamma^2(k,k') \psi_\alpha(k')
    \, 4\PI k'^2 \dd k'/ (2\PI)^3
\ee
[note $\gamma^2(k,k')$ here is not $[\gamma(k,k')]^2$ but rather
$\gamma^2(k,k') =
\int_0^\infty \gamma(k,k'') \gamma(k'',k') 4\PI k''^2 \dd k''/ (2\PI)^3$].
The corresponding minimum variances estimators $\hat\xi(k)$
and $\hat\xi_\alpha$ are similarly related,
by an expression which is essentially identical to~(\ref{xik}).

Being real and symmetric,
the scaling matrix $\gamma(k,k')$ must be diagonal
in some representation or other.
All the scaling matrices considered in the present paper
are diagonal in Fourier space,
so that the scaling matrix becomes a scaling function $\gamma(k)$.
In that case, the relation~(\ref{phi})
between the eigenfunctions $\phi_\alpha(k)$
and the kernels $\psi_\alpha(k)$ simplifies to
\be
\label{phis}
  \phi_\alpha(k) =
    \gamma(k) c_\alpha \psi_\alpha(k)
  \ ,
\ee
the orthogonality condition~(\ref{phiaphib}) becomes
\be
  \int_0^\infty \psi_\alpha(k) \psi_\beta(k) \gamma(k)^2
    \, {4\PI k^2 \dd k \over (2\PI)^3}
    =
  \left\{ \begin{array}{ll}
    0 & (\alpha \neq \beta) \\
    c_\alpha^{-2} & (\alpha = \beta)
  \end{array} \right.
  \ ,
\ee
and the expansion~(\ref{xik}) of the power spectrum $\xi(k)$
in the kernels $\psi_\alpha(k)$ reduces to
\be
  \xi(k) = \sum_\alpha \xi_\alpha c_\alpha^2 \psi_\alpha(k) \gamma(k)^2
  \ .
\ee
The eigenfunctions $\phi_\alpha(k)$ of the scaled Fisher matrix
$\gamma T \gamma$ are found by solving the eigenvalue equation~(\ref{eigeq}),
which reduces to
\be
\label{eigeqs}
  \int_0^\infty \gamma(k) T(k,k')
    \gamma(k') \phi_\alpha(k')
    \, {4\PI k' \dd k' \over (2\PI)^3}
    = \mu_\alpha \phi_\alpha(k)
  \ .
\ee
The relation~(\ref{eigvalT}) between the variances
$\langle \Delta\hat\xi_\alpha^2 \rangle$
and the eigenvalues $\mu_\alpha$ of the scaled Fisher matrix remains unchanged.

\section{Computation of the Fisher Matrix}
\label{compute}

A perturbation procedure for computing the Fisher matrix $T(k,k')$ of the
unredshifted power spectrum of a galaxy survey has been described in Paper~1.
Here I assume for simplicity Gaussian fluctuations,
and the zeroth order approximation,
which is essentially the classical approximation
considered by Feldman, Kaiser \& Peacock (1994, hereafter FKP),
where position and wavelength are simultaneously measurable.

To illustrate the procedures proposed in this paper,
I take the geometry of the PSCz survey,
which is the redshift survey of IRAS galaxies complete to 0.6~Jy
(Saunders et al., in preparation).
I take the angular geometry of PSCz to be the same as that of the QDOT survey
(Lawrence et al.~1997),
which covers most of the sky above galactic latitude $|b| > 10^\circ$.
I take the radial selection function to be 6 times that of QDOT,
which was designed to be a 1-in-6 random sampling of IRAS galaxies to 0.6~Jy.
The QDOT radial selection function is measured using Turner's (1979) method,
after making cosmological corrections for the volume element
(assuming a flat Universe),
and for evolution of the luminosity function
(adopting luminosity evolution $L \propto (1+z)^{1.8}$).
I chose also to exclude the local region closer than $25 \, h^{-1} \Mpc$,
to eliminate possible bias from the local overdensity;
in practice this has little effect at the long wavelengths
illustrated in this paper.

One test of the accuracy of the zeroth order approximation to the
Fisher matrix adopted here
is that the Fisher matrix should be strictly positive definite,
that is, all its eigenvalues should be strictly positive.
In the PSCz survey, I find that if the approximate Fisher matrix is computed
with a resolution in Fourier space of
$\Delta k = \PI/(896 \, h^{-1}\Mpc) = 0.00351 \, h\Mpc^{-1}$,
or coarser,
then all the eigenvalues are positive.
At finer resolutions an increasing fraction of the eigenvalues are negative.
For the trial bases exhibited in \S\ref{trial},
I chose to over-resolve the kernels, using
$\Delta k = \PI/(2048 \, h^{-1} \Mpc) = 0.00153 \, h \Mpc^{-1}$,
which appears to do no harm to kernels of low variance.
At this resolution $\sim 22 \%$ of the eigenvalues of the Fisher matrix
are negative,
but the absolute values of the negative eigenvalues are all less than
$10^{-4}$ of the largest eigenvalue.

The zeroth order approximation to the Fisher matrix is given
in Paper 1, equation~(86), as
\be
\label{Tabk1}
  T(k,k') =
    {1 \over 2} \int \bigl| U(\k+\k') \bigr|^2
    \, \dd o \dd o' / (4\PI)^2
\ee
where $\dd o$ denotes an interval of solid angle and the integration is over
all directions of $\k$ and $\k'$,
and where
$U(\k) = \int U(\r) \e^{\im \k.\r} \ddd r$
is the Fourier transform of
the FKP window
(Paper~1, eq.~[72])
\be
\label{U}
  U(\r) = {\Phi(\r) \over 1 + \xi(k_0) \Phi(\r)}
  \ .
\ee
Here $\Phi(\r)$ is the selection function,
the expected number density of galaxies at position $\r$ in the survey,
and $\xi(k_0)$ is the assumed prior power spectrum at wavenumber $k_0$.
The higher order terms of the Fisher matrix $T(k,k')$
comprise an expansion in $\xi(k)-\xi(k_0)$
(Paper~1).
Thus the greatest accuracy in the zeroth order term is accomplished by
choosing $k_0$ close to $k$ and $k'$;
I adopt
\be
\label{k0}
  k_0 = (k+k')/2
  \ .
\ee
The approximation should be quite good,
since the Fisher matrix~(\ref{Tabk1}) is almost diagonal in Fourier space,
its largest elements satisfying $k \approx k'$.

The power spectrum $\xi(k_0)$ in the survey window~(\ref{U})
can be regarded as a prior in Bayesian model testing,
or alternatively as a reasonable estimate which may in practice lead to kernels
yielding nearly statistically orthogonal sets of windowed power spectra.
I adopt the unredshifted power spectrum proposed by Peacock (1997).
Peacock gives analytic formulae for the power spectrum in two cases,
a flat model with $\Omega = 1$,
and an open model with $\Omega = 0.3$;
the results shown in the present paper are for the flat model.

\begin{figure}
\vbox to84mm{\rule{0pt}{84mm}}
\includegraphics{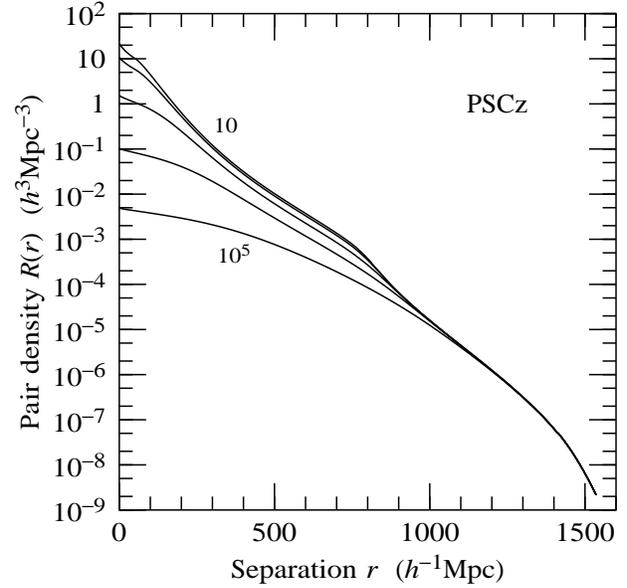}
  \caption[1]{
Expected density $R(r)$ of weighted random pairs at separation $r$,
equation~(\protect\ref{R}),
in the PSCz survey.
Curves, from top to bottom,
are for prior powers
$\xi(k_0) = 10$, $10^2$, $10^3$, $10^4$, and $10^5 \, h^{-3}\Mpc^3$
in the survey window $U(\r)$, equation~(\protect\ref{U}).
\label{winr}
}
\end{figure}

It is convenient to evaluate the Fisher matrix~(\ref{Tabk1}) via an
excursion back into real space.
In the real representation,
the Fisher matrix~(\ref{Tabk1}) is
\be
\label{Tabr}
  T(r,r') =
    \frac{1}{2} \delta_D(r - r') R(r)
\ee
where $R(r)$ is the integral of the pair window $U(\r_i) U(\r_j)$
over all pairs $ij$ in the survey separated by $r$
\be
\label{R}
  R(r) =
    \int U(\r_i) U(\r_j) \delta_D(|\r_i-\r_j|-r)
    \, \ddd r_i \ddd r_j
  \ .
\ee
Beware of equation~(\ref{Tabr})!
It does not signify that the Fisher matrix is diagonal in real space,
because the prior power spectrum $\xi(k_0)$ in the survey window $U(\r)$,
equation~(\ref{U}),
is a different constant for different elements of the Fisher matrix $T(k,k')$
(the Fisher matrix would be diagonal in real space if
the prior power spectrum were that of shot noise, $\xi(k)$ = constant,
but that is not the case here).
In accordance with the convention of Paper~1,
the Dirac delta-functions in equations~(\ref{Tabr}) and (\ref{R})
are normalized on a 3-dimensional scale, so that
$\int \delta_D(r-r') 4\PI r'^2 \dd r' = 1$.
The pair density $R(r)$,
which in the literature is often denoted $\langle R R \rangle$
(for random-random),
is the expected number of weighted random pairs of galaxies in the survey
per unit volume interval $4\PI r^2 \dd r$ at separation $r$.
Although the pair density is commonly computed by Monte Carlo integration,
I find it faster, more accurate,
and more convenient (since the program was already written)
to compute the integral directly,
using the procedures described by
Hamilton (1993).
Figure~\ref{winr} illustrates the pair density $R(r)$ in the PSCz survey
for various values of the prior power $\xi(k_0)$
upon which the survey window $U(\r)$ depends, equation~(\ref{U}).

\begin{figure}
\vbox to84mm{\rule{0pt}{84mm}}
\includegraphics{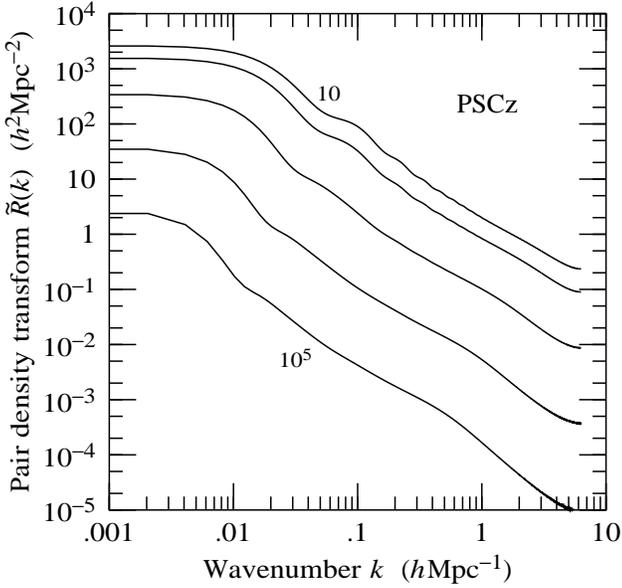}
  \caption[1]{
One-dimensional Fourier transform $\tildeR(k)$,
equation~(\protect\ref{Rk}),
of the pair density $R(r)$, Figure~\protect\ref{winr}, in the PSCz survey.
Curves, from top to bottom,
are for prior powers
$\xi(k_0) = 10$, $10^2$, $10^3$, $10^4$, and $10^5 \, h^{-3}\Mpc^3$.
\label{wink}
}
\end{figure}

Transforming the Fisher matrix~(\ref{Tabr}) from real space back into
Fourier space,
$
  T(k,k') =
    \int T(r,r')
    \discretionary{}{}{}
    \e^{\im \k . \r + \im \k' . \r'}
    \discretionary{}{}{}
    \ddd r \ddd r'
$,
yields
\be
\label{Tabk}
  T(k,k') =
    {\PI \over 2 k k'}
    \bigl[ \tildeR(k-k') - \tildeR(k+k') \bigr]
\ee
where $\tildeR(k)$ is the 1-dimensional Fourier transform of the pair density
$R(r)$
\be
\label{Rk}
  \tildeR(k) \equiv 2 \int_0^\infty R(r) \cos(kr) \, \dd r
  \ .
\ee
Figure~\ref{wink} illustrates the pair density transform $\tildeR(k)$
in the PSCz survey.
As expected, the pair density transform is peaked to small wavenumber $k$.

The resolution in $k$-space with which the power spectrum can be measured
is roughly set by the inverse scale of the survey,
so it makes sense to discretize the Fisher matrix $T(k_m,k_n)$
on a uniform grid in $k$.
A uniform grid in $k$ is also a natural choice numerically,
both in evaluating the pair density transform $\tildeR(k)$,
equation~(\ref{Rk}),
by fast Fourier transform,
and then in constructing the Fisher matrix~(\ref{Tabk})
from $\tildeR(k)$ evaluated at $k = k_m \pm k_n$.

I proceed numerically as follows.
Firstly, compute the pair density $R(r)$,
using the method detailed by Hamilton (1993),
over a table of separations $r$ and constants $\xi(k_0)$,
logarithmically spaced at $0.02$ dex in $r$,
and $0.1$ dex in $\xi(k_0)$.
To facilitate taking a fast Fourier transform,
interpolate on $R(r)$ to define a new table linearly spaced
typically at intervals $0.5 \, h^{-1}\Mpc$ in $r$.
Optionally zero-pad this table,
typically to a separation of $2048 \, h^{-1}\Mpc$,
which results in a grid spacing in $k$-space of
$\Delta k = \PI/(2048 \, h^{-1} \Mpc) = 0.00153 \, h \Mpc^{-1}$.
Although zero-padding increases the resolution
with which the Fisher matrix and its eigenfunctions are sampled,
it does not increase the physical resolution available from the survey.
For each matrix element $T(k_m,k_n)$,
choose $k_0 = (k_m + k_n)/2$,
equation~(\ref{k0});
set $\xi(k_0)$ to be the $\Omega = 1$ power spectrum proposed by Peacock (1997);
interpolate logarithmically on the table of the pair density $R(r)$
as a function of $\xi(k_0)$ to find $R(r)$ for the desired $\xi(k_0)$;
fast Fourier (cosine) transform $R(r)$ to find $\tildeR(k)$,
equation~(\ref{Rk});
and thence obtain $T(k_m,k_n)$,
equation~(\ref{Tabk}).
Since $k_0 = (k_m + k_n)/2$ is constant along cross-diagonals
of the Fisher matrix $T(k_m,k_n)$,
the fast Fourier transform need be carried out only for each cross-diagonal,
not for each element of the matrix.

The above procedure yields the Fisher matrix $T(k_m,k_n)$
on a uniformly spaced grid of $k_n = n \Delta k$
with $n = 0,1,\ldots,N$.
According to \S\ref{orthog},
bases of kernels yielding statistically orthogonal sets of windowed power
spectra are constructed by diagonalizing matrices of the form
$\gamma T \gamma$,
where $\gamma$ is some strictly positive definite symmetric matrix.
In all cases considered in the present paper the scaling matrix $\gamma$
is diagonal in Fourier space,
that is, $\gamma(k)$ is some strictly positive function of $k$.
The equation to be solved is then the eigenvalue equation~(\ref{eigeqs}).
In discretized form, this eigenvalue equation~(\ref{eigeqs}) becomes
(there is no implicit summation in the following equation,
nor elsewhere in this paper)
\be
\label{eigeqF}
  \sum_{n = 0}^N \gamma_m F_{m n} \gamma_n u_{n\alpha} = \mu_\alpha u_{m\alpha}
\ee
where $\gamma_m \equiv \gamma(k_m)$,
the matrix $F_{mn}$ is a discrete version of the Fisher matrix $T$
\be
\label{F}
  F_{m n} \equiv
    k_m T(k_m,k_n) k_n
    \, 4\PI \Delta k / (2\PI)^3
  \ ,
\ee
the eigenvalues $\mu_\alpha$ are the same as in equation~(\ref{eigeqs}),
and the eigenfunctions $u_{m\alpha} \equiv u_\alpha(k_m)$,
satisfying the orthonormality condition
$\sum_{m = 0}^N u_\alpha(k_m) u_\beta(k_m) = \delta_{\alpha \beta}$,
are discrete versions of the eigenfunctions $\phi_\alpha(k_m)$
\be
\label{u}
  u_{m\alpha} \equiv u_\alpha(k_m) \equiv
    k_m \phi_\alpha(k_m) \left[ 4\PI \Delta k / (2\PI)^3 \right]^{1/2}
  \ .
\ee
It is useful also to define corresponding discrete versions
$v_{m\alpha} \equiv v_\alpha(k_m)$
of the kernels $\psi_\alpha(k_m)$
\be
\label{v}
  v_{m\alpha} \equiv v_\alpha(k_m) \equiv
    k_m \psi_\alpha(k_m) \left[ 4\PI \Delta k / (2\PI)^3 \right]^{1/2}
\ee
which are related to the discrete eigenfunctions $u_\alpha(k_m)$
by the same formula~(\ref{phis}) which relates the kernels
$\psi_\alpha(k_m)$ and the eigenfunctions $\phi_\alpha(k_m)$,
with the same normalization constants $c_\alpha$,
\be
\label{us}
  u_\alpha(k_m) =
    \gamma(k_m) c_\alpha v_\alpha(k_m)
  \ .
\ee
If the kernels $\psi_\alpha(k)$ are normalized on the quadratic
convention~(\ref{norm}),
then the discrete kernels $v_\alpha(k_m)$ are similarly quadratically
normalized,
$\sum_{m=0}^N v_\alpha(k_m)^2 = 1$.
In terms of the discrete kernel $v_\alpha(k_m)$, the windowed power spectrum
$\xi_\alpha$, equation~(\ref{xia}), is
\be
\label{xiav}
  \xi_\alpha = \sum_{m=0}^{N} v_\alpha(k_m) \xi(k_m)
    k_m \left[ 4\PI \Delta k / (2\PI)^3 \right]^{1/2}
  \ .
\ee

Although the indices $m$ and $n$ in the eigenvalue equation~(\ref{eigeqF})
run over values 0 to $N$,
it follows from equation~(\ref{Tabk}),
and the periodicity of the discrete cosine transform $\tildeR(k_n)$
over the interval $n = [-N,N]$,
that the perimeter rows and columns of $F$ are identically zero,
$F_{m n} = 0$ for $m$ or $n$ = 0 or $N$.
Amongst other things,
this guarantees that each eigenfunction $u_\alpha(k_m)$,
and consequently each kernel $v_\alpha(k_m)$,
vanishes at the boundary values $m = 0$ and $N$.
Thus $F_{m n}$ is effectively reduced to a $(N-1) \times (N-1)$ matrix,
with $m$ and $n$ running over values 1 to $N-1$.

Solution of equation~(\ref{eigeqF})
yields $N-1$ orthonormal eigenvectors $u_\alpha(k_m)$,
and the kernels $v_\alpha(k_m)$ then follow from equation~(\ref{us}).
The eigenfunctions $\phi_\alpha(k_m)$ and the corresponding kernels
$\psi_\alpha(k_m)$ are given in terms of their discrete counterparts
$u_\alpha(k_m)$ and $v_\alpha(k_m)$ by equations~(\ref{u}) and (\ref{v}).
The expected variance
$\langle \Delta\hat\xi_\alpha^2 \rangle$
of estimates $\hat\xi_\alpha$ of the power spectrum
windowed through the kernel $\psi_\alpha$,
equations~(\ref{xia}) and (\ref{xihata}),
is related to the eigenvalue $\mu_\alpha$ of $\gamma F \gamma$
in equation~(\ref{eigeqF}) by equation~(\ref{eigvalT}).

In practice,
I do not attempt to solve for the eigenfunctions of the full matrix
$\gamma F \gamma$, but truncate the matrix typically
at a half-wavelength of $\PI/(N\Delta k) = 8 \, h^{-1}\Mpc$.
At a resolution of
$\Delta k = \PI/(2048 \, h^{-1} \Mpc)$,
this means truncating the matrix to the first $N-1=255$ rows and columns
(i.e., I only bother to calculate these elements).
The cutoff at $8 \, h^{-1}\Mpc$ seems reasonable,
given that the Fisher matrix calculated here assumes Gaussian fluctuations,
which may be valid on linear scales, down to $\sim 8 \, h^{-1}\Mpc$,
but certainly fails below this.

\section{Some Trial Kernels}
\label{trial}

As discussed in \S\ref{orthog},
there are many distinct bases of kernels $\psi_\alpha(k)$ which yield
statistically orthogonal sets of windowed power spectra
$\hat\xi_\alpha$.
Any such basis can be constructed by choosing a strictly positive definite
scaling matrix $\gamma$,
and finding the eigenfunctions $\phi_\alpha = \gamma c_\alpha \psi_\alpha$
of the scaled Fisher matrix $\gamma T \gamma$.
In all the cases considered here,
the scaling matrix $\gamma$ is taken to be diagonal in Fourier space,
so that it becomes a scaling function $\gamma(k)$.

In this section I show, for the PSCz geometry,
bases of kernels for two natural choices of the scaling function,
$\gamma(k) = 1$ in \S\ref{raw},
and $\gamma(k) = \xi(k)$ in \S\ref{ln}.

As will be seen, the resulting kernels are too broad and wiggly in
$k$-space to be of much practical use.
The difficulty is resolved in the next Section, \S\ref{compact},
where a basis of kernels which are narrow and positive in Fourier space
will be constructed.

\subsection{Eigenfunctions of
$\langle \Delta\hat\xi(k) \Delta\hat\xi(k') \rangle$}
\label{raw}

\begin{figure}
\vbox to84mm{\rule{0pt}{84mm}}
\includegraphics{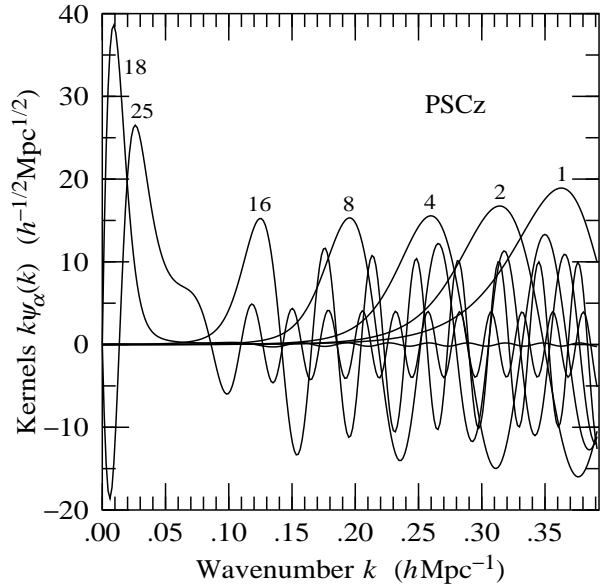}
  \caption[1]{
A sample of eigenfunctions $\psi_\alpha(k)$ of the expected covariance
$\langle \Delta\hat\xi(k) \Delta\hat\xi(k') \rangle$
of minimum variance estimates of the power spectrum of the PSCz survey.
The eigenfunctions are labelled in order of their eigenvalues,
the expected variances
$\langle \Delta\hat\xi_\alpha^2 \rangle$,
the lowest numbered eigenfunction having the smallest variance.
The grid in $k$ here is $N=256$ long,
so there are $N-1=255$ eigenmodes.
The resolution is
$\Delta k = \PI/(2048 \, h^{-1} \Mpc) = 0.00153 \, h \Mpc^{-1}$.
\label{rawfish}
}
\end{figure}

A natural choice of the scaling function $\gamma(k)$ is a constant
\be
\label{gam1}
  \gamma(k) = 1
  \ .
\ee
The kernels $\psi_\alpha(k)$ in this case
coincide with the scaled kernels $\phi_\alpha(k)$,
and are the eigenfunctions of the covariance matrix
$\langle \Delta\hat\xi(k) \Delta\hat\xi(k') \rangle$.
As discussed in \S\ref{compute},
these eigenfunctions are in practice constructed as eigenfunctions
of the Fisher matrix
$T(k,k') = \langle \Delta\hat\xi(k) \Delta\hat\xi(k') \rangle^{-1}$.
The kernels $\psi_\alpha(k)$ here form an orthonormal set,
a property unique to the case of a constant scaling function.

Figure~\ref{rawfish}
shows a selection of eigenfunctions $\psi_\alpha(k)$
of the Fisher matrix $T(k,k')$ of the PSCz survey.
The Figure shows $k \psi_\alpha(k)$ rather than $\psi_\alpha(k)$
in part because the discretized version $v_\alpha(k)$ of the kernel,
equation~(\ref{v}),
includes the extra factor of $k$,
and in part because $k \psi_\alpha(k)$
gives a better impression of the contribution
to the windowed power spectrum $\xi_\alpha$ from the power $\xi(k)$
at different $k$,
given that the kernel $\psi_\alpha(k)$ gets multiplied by $k^2 \xi(k)$
and that $\xi(k) \simpropto k^{-1}$
is a reasonable intermediate choice for the slope of the power spectrum.

The kernels shown in Figure~\ref{rawfish}
have the drawbacks that for the most part
they are neither everywhere positive,
nor compact in $k$.
A third problem is that when the range in $k$ is extended to shorter
wavelengths (larger $k$),
the fundamental mode at the shortest scales retreats
to the smallest available scale.
These defects are redressed in \S\ref{compact}.

The kernels shown in Figure~\ref{rawfish}
exhibit some features which are typical also
for other choices of the scaling function $\gamma(k)$.
The first several (here about 30) modes cover the full range of $k$
with relatively broad kernels of relatively low variance.
Subsequent modes (not shown) retrace the same range with ever
more oscillatory kernels.
The shapes of the noisiest kernels resemble
violin notes viewed on an oscilloscope.
Unlike the fundamental at small scales (large $k$),
the fundamental at large scales (small $k$)
appears robust against changes of resolution in either real or Fourier space,
as long as the resolution in Fourier space is high enough.

The large scale fundamental, which is almost but not quite positive everywhere,
happens to have the 18th smallest variance
$\langle \Delta\hat\xi_\alpha^2 \rangle$.
One should bear in mind that the variance depends on the normalization
of the kernel, which is taken here to be quadratic, equation~(\ref{norm}).
A linear normalization would tend to increase the amplitude
and hence variance of the wigglier kernels,
and would improve the rating of the fundamental.
However, the ordering of the variances is correct for the
kernels as normalized in Figure~\ref{rawfish}.

\subsection{Eigenfunctions of
$\xi(k)^{-1} \! \langle \Delta\hat\xi(k) \Delta\hat\xi(k') \rangle \xi(k')^{-1}$
}
\label{ln}

\begin{figure}
\vbox to84mm{\rule{0pt}{84mm}}
\includegraphics{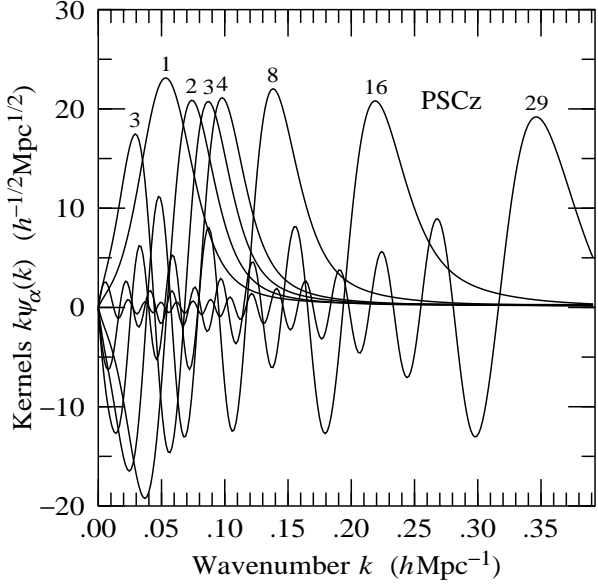}
  \caption[1]{
A sample of kernels $\psi_\alpha(k) \propto \phi_\alpha(k)/\xi(k)$
obtained from the eigenfunctions $\phi_\alpha(k)$
of the expected relative covariance
$\xi(k)^{-1} \! \langle \Delta\hat\xi(k) \Delta\hat\xi(k') \rangle \xi(k')^{-1}$
of minimum variance estimates of the power spectrum of the PSCz survey.
The eigenfunctions are labelled in order of the eigenvalues,
the relative variances
$\langle (\Delta\hat\xi/\xi)^2_\alpha \rangle$,
equation~(\protect\ref{Dxirel}),
the lowest eigenfunction having the smallest relative variance.
The fundamental mode is centred about the peak in the prior power spectrum
at $k \approx 0.05 \, h\Mpc^{-1}$.
\label{lnfish}
}
\end{figure}

For a contiguous survey of extremely large (tending to infinity) volume $V$,
and for Gaussian fluctuations,
the power spectrum itself becomes statistically orthogonal,
with variance
(Paper 1, eq.~[65] with [63])
\be
\label{DxikDxik}
  \langle \Delta\hat\xi(k) \Delta\hat\xi(k') \rangle =
    (2\PI)^3 \delta_D(k-k')
    {2 \xi(k)^2 \over V}
\ee
where the delta-function $(2\PI)^3 \delta_D(k-k')$
is the unit matrix in the Fourier representation.
It follows from equation~(\ref{DxikDxik})
that the relative covariance
$\xi^{-1} \! \langle \Delta\hat\xi \Delta\hat\xi \rangle \xi^{-1}$
of the power spectrum
in an extremely large survey is proportional to the unit matrix
\be
\label{DlnxikDlnxik}
  {\langle \Delta\hat\xi(k) \Delta\hat\xi(k') \rangle \over \xi(k) \xi(k')} =
    (2\PI)^3 \delta_D(k-k')
    {2 \over V}
  \ .
\ee

This suggests that it might be interesting to consider kernels $\psi_\alpha(k)$
derived from the eigenfunctions $\phi_\alpha(k)$
of the relative covariance matrix
$\xi(k)^{-1} \!
\langle \Delta\hat\xi(k) \Delta\hat\xi(k') \rangle \xi(k')^{-1}$,
corresponding to the case where the scaling function $\gamma(k)$
is chosen equal to the prior power spectrum $\xi(k)$
\be
  \gamma(k) = \xi(k)
  \ .
\ee
The eigenvalues $\lambda_\alpha$ of the relative covariance matrix
$\xi(k)^{-1} \! \langle \Delta\hat\xi(k) \Delta\hat\xi(k') \rangle \xi(k')^{-1}$
are equal to the relative variances
$\langle (\Delta\hat\xi/\xi)_\alpha^2 \rangle$
\be
  \lambda_\alpha =
    \langle (\Delta\hat\xi/\xi)_\alpha^2 \rangle
\ee
defined by
\be
\label{Dxirel}
  \langle (\Delta\hat\xi/\xi)_\alpha^2 \rangle \equiv
    {\langle \Delta\hat\xi_\alpha^2 \rangle \over
    \int_0^\infty \bigl[ \psi_\alpha(k) \xi(k) \bigr]^2
    \, 4\PI k^2 \dd k / (2\PI)^3}
  \ .
\ee
The relative variances
$\langle (\Delta\hat\xi/\xi)_\alpha^2 \rangle$,
which can be regarded as being defined for arbitrary kernels $\psi_\alpha(k)$
by equation~(\ref{Dxirel}),
have units of inverse volume,
and are equal to $2/V$ in the limit of a very large volume,
according to equation~(\ref{DlnxikDlnxik}).
It may plausibly be argued that the relative variances
$\langle (\Delta\hat\xi/\xi)_\alpha^2 \rangle$
may offer a better measure of noisiness than the absolute variances
$\langle \Delta\hat\xi_\alpha^2 \rangle$,
since equation~(\ref{DxikDxik}) shows that the variance of
the power spectrum windowed through some kernel $\psi_\alpha(k)$
may be large not because of noise,
but simply because $\xi(k)$ itself is large over the range
of the kernel $\psi_\alpha(k)$.

Figure~\ref{lnfish} shows the resulting kernels
$\psi_\alpha(k) \propto \phi_\alpha(k)/\xi(k)$ derived from the eigenfunctions
$\phi_\alpha(k)$ of the relative covariance matrix
$\xi(k)^{-1} \!
\langle \Delta\hat\xi(k) \Delta\hat\xi(k') \rangle \xi(k')^{-1}$.
As usual, the eigenfunctions are in practice found as eigenfunctions of
the inverse of this matrix, the scaled Fisher matrix
$\xi(k) T(k,k') \xi(k')$.

The kernels are labelled not in order of the absolute variances
$\langle \Delta\hat\xi_\alpha^2 \rangle$ as in Figure~\ref{rawfish},
but rather in order of the eigenvalues, the relative variances
$\langle (\Delta\hat\xi/\xi)_\alpha^2 \rangle$,
equation~(\ref{Dxirel}).
The single fundamental mode has the smallest relative variance,
and is centred around the maximum at
$\sim 0.05 \, h\Mpc^{-1}$ in the prior power spectrum.

The kernels shown in Figure~\ref{lnfish} share the same problem
as those in Figure~\ref{rawfish},
that they are neither compact nor everywhere positive in Fourier space.

\section{A Basis of Positive Compact Kernels}
\label{compact}

It is evidently desirable that a basis of kernels,
besides yielding a statistically orthogonal set of windowed power spectra,
should be positive and compact in Fourier space.
Positivity would preserve
the intrinsic positivity of the power spectrum,
while compactness would allow the power spectrum to be measured
with high resolution in Fourier space.
In this Section I show how to construct such a basis.
A numerically convenient version of the construction is described in
the paragraph containing equations~(\ref{vf}) and (\ref{Dxif}).

To understand how to make kernels compact and positive,
consider the mathematical theorem that the eigenmodes of a real symmetric matrix
are unique up to arbitrary orthogonal rotations amongst degenerate eigenmodes,
that is, amongst eigenmodes with the same eigenvalue.
In the present case the eigenvalues of the Fisher matrix,
even if not degenerate,
are nevertheless many and finely spaced,
and there is much potential for nearly degenerate eigenmodes to
mix in random unpleasant ways.
This suggests that mixing can be reduced in Fourier space
by lifting the degeneracy of eigenvalues in Fourier space,
which can be accomplished by applying a
scaling function $\gamma(k)$
which is a strongly varying function of $k$.
Numerical experiment indicates that
the most compact and positive kernels arise
from choosing the scaling function $\gamma(k)$ to be as steep as possible.

The limiting case of an infinitely steep scaling function $\gamma(k)$
can be solved analytically,
and as will be seen, Figure~\ref{goodfish}, it gives rise to a basis
of positive kernels as pretty as one could wish for.
The solution, given immediately following,
is simpler than for finite scaling functions $\gamma(k)$,
requiring no diagonalization at all.

Suppose that the matrix $T(k_m,k_n)$ is discretized,
as is the practical case, with $m$, $n = 1, 2, \dots, N\!-\!1$,
and suppose that $\gamma(k_1) \gg \gamma(k_2) \gg \ldots \gg \gamma(k_{N-1})$.
For a uniformly spaced grid of $k$, as here,
it is convenient to work with the discrete version $F_{mn}$,
equation~(\ref{F}),
of the Fisher matrix,
and the associated discrete versions $u_\alpha(k_m)$ and $v_\alpha(k_m)$
of the eigenfunctions $\phi_\alpha(k_m)$
and the kernels $\psi_\alpha(k_m)$,
equations~(\ref{u}) and (\ref{v}).
The equation to be solved is the discrete eigenvalue equation~(\ref{eigeqF}).

Since the first column of the scaled Fisher matrix
$\gamma_m F_{mn} \gamma_n$
(no implicit summation)
is huge compared to all other columns,
the first eigenfunction $u_1(k_m)$,
the one with largest eigenvalue,
is proportional to this first column,
$u_1(k_m) \propto \gamma_m F_{m1}$.
It follows that the first kernel $v_1(k_m) \propto u_1(k_m) / \gamma_m$
is proportional to the first column of the Fisher matrix,
$v_1(k_m) \propto F_{m1}$.
Since the first element of the eigenvector $u_1(k_m)$ is huge
compared to all other elements,
the eigenvector is $u_1(k_m) = (1,0,\ldots,0)$ in the limit,
but the higher elements of the kernel vector $v_1(k_m)$
remain generally nonzero.

The remaining eigenvectors $u_\alpha(k_m)$, $\alpha = 2,\ldots,N-1$,
must be orthogonal to the first eigenvector $u_1(k_m)$,
which means that their first elements must be zero,
$u_\alpha(k_1) = 0$ for $\alpha > 1$.
Since the first two columns of the scaled Fisher matrix
are huge compared to the others,
the second eigenfunction $u_2(k_m)$, with the second largest eigenvalue,
must be some linear combination of the first two columns
$\gamma_m F_{m1}$ and $\gamma_m F_{m2}$,
and it must be proportional to that linear combination whose first element
vanishes,
$u_2(k_m) \propto \gamma_m ( F_{11} F_{m2} - F_{12} F_{m1} )$.
The second kernel follows from
$v_2(k_m) \propto u_2(k_m)/\gamma_m \propto F_{11} F_{m2} - F_{12} F_{m1}$,
whose first element is zero but whose other elements are finite.
The second eigenfunction is $u_2(k_m) = (0,1,0,\ldots,0)$ in the limit.
Proceeding, one concludes that the $\alpha$th eigenfunction $u_\alpha(k_m)$
is that linear combination of the first $\alpha$ columns of the scaled
Fisher matrix which is zero in the first $\alpha-1$ elements.
Normalized to $u_\alpha(k_\alpha) = 1$, the eigenfunction is
\be
\label{uc}
  u_\alpha(k_m) = \left\{ \begin{array}{ll}
    \displaystyle
    {\gamma_\alpha F_{m\alpha}^{-1} \over \gamma_m F_{\alpha\alpha}^{-1}}
      & (m \le \alpha) \\
    \displaystyle
    {\gamma_m \over \gamma_\alpha} \sum_{n=1}^\alpha F_{mn} F_{n\alpha}^{-1}
      & (m \ge \alpha)
  \end{array} \right.
\ee
where $F_{m\alpha}^{-1}$ denotes the $(m,\alpha)$th element of the
inverse of the $\alpha \times \alpha$ submatrix consisting of the first
$\alpha$ rows and columns of $F_{mn}$.
In the limit, the eigenfunction $u_\alpha(k_m)$
is $1$ in the $\alpha$th place and zero in all others:
\be
\label{uclim}
  u_\alpha(k_m) \rightarrow \left\{ \begin{array}{ll}
    1 & (m = \alpha) \\
    0 & (m \neq \alpha) \ .
  \end{array} \right.
\ee
The associated eigenvalue of the scaled Fisher matrix is
\be
\label{muc}
  \mu_\alpha = {\gamma_\alpha^2 \over F_{\alpha\alpha}^{-1}}
  \ .
\ee

It follows from equation~(\ref{uc})
that the kernel $v_\alpha(k_m) \propto u_\alpha(k_m)/\gamma_m$
is given by
\be
\label{vc}
  v_\alpha(k_m) =
    c'^{-1}_\alpha \sum_{n=1}^\alpha F_{mn} F_{n\alpha}^{-1}
\ee
where
$c'_\alpha$
is an arbitrary normalization constant.
Note $v_\alpha(k_m) = 0$ for $m \leq \alpha - 1$.
For the quadratic normalization convention adopted in this paper,
$\sum_{m=1}^{N-1} v_\alpha(k_m)^2 = 1$,
the normalization constant is
$c'_\alpha = \left[ \sum_{m=1}^{N-1}
\left( \sum_{n=1}^\alpha F_{mn} F_{n\alpha}^{-1} \right)^2 \right]^{1/2}$.
The usual normalization constants $c_\alpha$
which relate $\phi_\alpha$ and $\psi_\alpha$, equation~(\ref{phis}),
or equivalently $u_\alpha$ and $v_\alpha$, equation~(\ref{us}),
are related to the normalization constants $c'_\alpha$ above by
\be
\label{cc}
  c_\alpha = c'_\alpha / \gamma_\alpha
  \ .
\ee
The estimators $\hat\xi_\alpha$,
equations~(\ref{xia}), (\ref{xihata}), and (\ref{xiav}),
of power spectra windowed through
the kernels $\psi_\alpha(k)$
corresponding to $v_\alpha(k)$, equation~(\ref{v}),
are by construction statistically orthogonal,
in accordance with equation~(\ref{DxiDxi}).
The expected variance
$\langle \Delta\hat\xi_\alpha^2 \rangle$
is given by formula~(\ref{eigvalT}), which here reduces to
\be
  \langle \Delta\hat\xi_\alpha^2 \rangle =
    {F_{\alpha\alpha}^{-1} \over c'^2_\alpha}
\ee
the factors of $\gamma_\alpha$ in the eigenvalue $\mu_\alpha$,
equation~(\ref{muc}),
and the normalization constant $c_\alpha$, equation~(\ref{cc}), cancelling out.

\begin{figure*}
\begin{minipage}{175mm}
\vbox to84mm{\rule{0pt}{84mm}}
\includegraphics{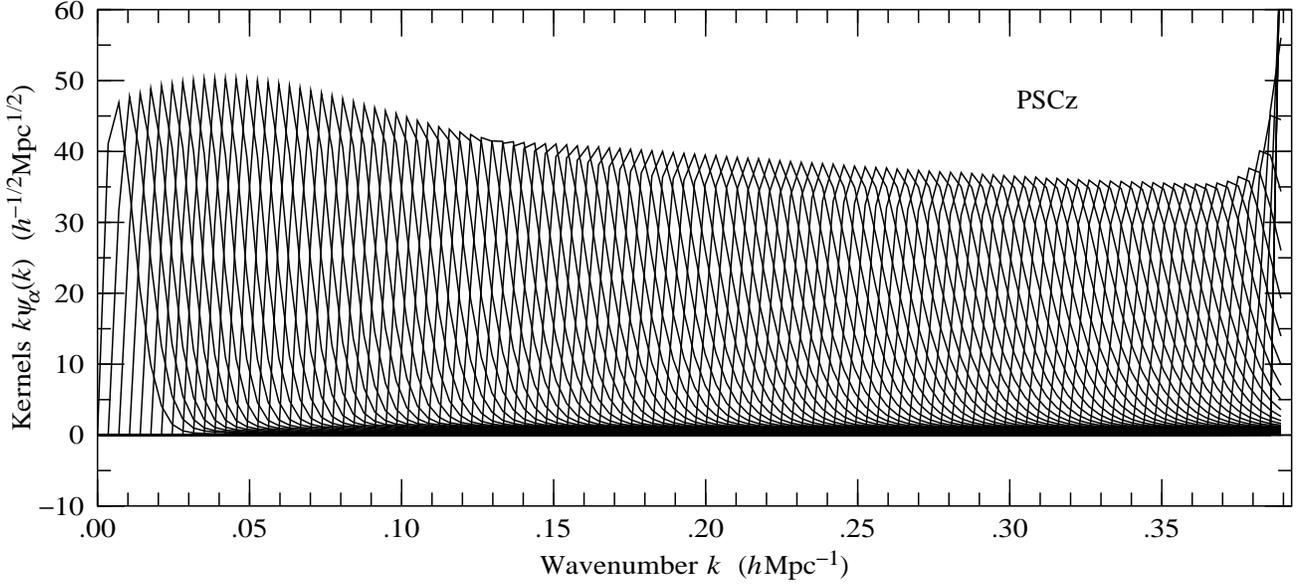}
  \caption[1]{
Basis of positive kernels obtained by applying an infinitely steep
scaling function to the Fisher matrix of the PSCz survey,
as described in the text, \S\protect\ref{compact}.
Counterintuitively, notwithstanding the overlap between kernels,
the minimum variance estimators $\hat\xi_\alpha$
of power spectra windowed through these kernels are
statistically orthogonal.
The resolution is
$\Delta k = \PI/(896 \, h^{-1}\Mpc) = 0.00351 \, h\Mpc^{-1}$,
which is the highest resolution such that all the eigenvalues are positive.
The grid in $k$ is $N=112$ long, so there are $N-1=111$ kernels.
The crowding of the kernels at the right of the graph is an artefact
of truncating the Fisher matrix at
$k_N = N\Delta k = \PI/(8 \, h^{-1}\Mpc) = 0.393 \, h\Mpc^{-1}$,
and could easily be avoided by extending the matrix to shorter wavelengths
(larger $k$).
\label{goodfish}
}
\end{minipage}
\end{figure*}

In practice,
the kernels $v_\alpha(k_m)$, equation~(\ref{vc}),
can be evaluated conveniently by the following procedure,
which reduces the Fisher matrix $F_{mn}$ iteratively to lower triangular form.
Suppose that the discrete Fisher matrix $F_{mn}$ defined by equation~(\ref{F})
has been reduced so far to the point where all the above diagonal elements
of the first $\alpha-1$ rows are zero.
Call this reduced Fisher matrix $F'_{mn}$,
which starts equal to the Fisher matrix $F_{mn}$ at step one, $\alpha = 1$.
Then the $\alpha$th kernel $v_\alpha(k_m)$ is proportional to the
$\alpha$th column of the reduced Fisher matrix
\be
\label{vf}
  v_\alpha(k_m) = {F'_{m\alpha} \over c''_\alpha}
\ee
where
$c''_\alpha$
is an arbitrary normalization constant.
For the quadratic normalization convention
$\sum_{m=1}^{N-1} v_\alpha(k_m)^2 = 1$,
the normalization constant is
$c''_\alpha = \left( \sum_{m=1}^{N-1} F'^2_{m\alpha} \right)^{1/2}$.
The previous normalization constant $c'_\alpha$ in equation~(\ref{vc})
is related to the normalization constant $c''_\alpha$ by
$c'_\alpha = c''_\alpha F_{\alpha\alpha}^{-1}$.
The quantity $F_{\alpha\alpha}^{-1}$,
which was defined immediately following equation~(\ref{uc}),
and upon which the eigenvalue $\mu_\alpha$, equation~(\ref{muc}),
also depends, is equal to the reciprocal of the diagonal element
of the reduced Fisher matrix,
$F_{\alpha\alpha}^{-1} = 1/F'_{\alpha\alpha}$.
Thus the eigenvalue~(\ref{muc}) is
$\mu_\alpha = \gamma_\alpha^2 F'_{\alpha\alpha}$.
The variance
$\langle \Delta\hat\xi_\alpha^2 \rangle$
of the windowed power spectrum $\hat\xi_\alpha$ is
\be
\label{Dxif}
  \langle \Delta\hat\xi_\alpha^2 \rangle =
    {F'_{\alpha\alpha} \over c''^2_\alpha}
  \ .
\ee
Now zero out the $\alpha$th row of $F'_{mn}$ above the diagonal
by subtracting an appropriate multiple of the $\alpha$th column
$F'_{m\alpha}$
from each higher column,
that is,
$F'_{mn} \rightarrow F'_{mn} - F'_{m\alpha} F'_{\alpha n} / F'_{\alpha\alpha}$
for $n \ge \alpha+1$.
Relabel the resulting matrix again $F'_{mn}$.
Repeat the above steps until the matrix is in lower triangular form.

When done,
the columns of the lower triangular matrix $F'_{m\alpha}$
will be proportional to the kernels $v_\alpha(k_m)$,
equation~(\ref{vf}),
while the diagonal elements $F'_{\alpha\alpha}$
will be proportional to the variances
$\langle \Delta\hat\xi_\alpha^2 \rangle$,
equation~(\ref{Dxif}).
Note that the first $\alpha-1$ elements of the $\alpha$th kernel are zero,
$v_\alpha(k_m) = 0$ for $m \le \alpha-1$;
in particular, the final kernel is $v_{N-1}(k_m) = (0,\dots,0,1)$.

If the Fisher matrix were strictly positive definite,
that is, if all its eigenvalues were strictly positive,
as should be the case theoretically,
then all the variances $\langle \Delta\hat\xi_\alpha^2 \rangle$,
equation~(\ref{Dxif}),
would be guaranteed to be positive.
As remarked near the beginning of \S\ref{compute},
the eigenvalues of the approximate Fisher matrix of the PSCz survey
computed here are all positive only if the resolution is taken to be
$\Delta k = \PI/(896 \, h^{-1}\Mpc) = 0.00351 \, h\Mpc^{-1}$
or coarser.
This, apparently, sets the natural resolution of the PSCz survey.

Figure~\ref{goodfish}
shows the resulting basis of kernels for the PSCz survey
at the `natural' resolution
$\Delta k = \PI/(896 \, h^{-1}\Mpc)$.
All of the $N-1=111$ eigenvalues down to the cutoff half-wavelength of
$\PI/(N\Delta k) = 8 \, h^{-1}\Mpc$ are positive,
and all the 111 kernels are everywhere positive.
At higher resolutions an increasing fraction of the eigenvalues
of the Fisher matrix are negative.
At a resolution of
$\Delta k = \PI/(1024 \, h^{-1}\Mpc) = 0.00307 \, h\Mpc^{-1}$
there are 3 negative of 127 eigenvalues,
while at
$\Delta k = \PI/(1536 \, h^{-1}\Mpc) = 0.00204 \, h\Mpc^{-1}$
there are 27 negative of 191 eigenvalues.
The kernels corresponding to negative eigenvalues are not positive everywhere,
and they are not well-behaved.
Kernels with positive eigenvalues remain positive
and mostly well-behaved,
even when kernels with negative eigenvalues are present,
although there is an increasing tendency for kernels,
especially those adjacent to kernels with negative eigenvalues,
to become shaky at finer resolutions.

It is evident that the kernels in Figure~\ref{goodfish} overlap substantially.
Yet they yield, by construction,
a statistically orthogonal set of windowed power spectra.
This seems astonishing: how can it be?
The answer is that the non-diagonal elements of the covariance
$\langle \Delta\hat\xi(k) \Delta\hat\xi(k') \rangle$
of the power spectrum of the survey are not all positive,
and the positive and negative covariances cancel
precisely between overlapping kernels.
Rather,
it is the elements of its inverse, the Fisher matrix
$T(k,k') = \langle \Delta\hat\xi(k) \Delta\hat\xi(k') \rangle^{-1}$,
which are all positive in Fourier space,
at least in the approximation used here, equation~(\ref{Tabk1}).
If the elements of the Fisher matrix are all positive,
then necessarily the elements of its inverse cannot all be positive.
Numerically, I find that the elements of the (discretized) covariance
$\langle \Delta\hat\xi(k) \Delta\hat\xi(k') \rangle$
in the PSCz survey oscillate wildly between positive and negative values.
This is not simply a numerical artefact.
In real space, the survey correlation function is a sum
$\xi(r_{ij}) + \Phi(\r_i)^{-1} \delta_D(\r_{ij})$
of a cosmic term $\xi(r_{ij})$ and a Poisson sampling term
$\Phi(\r_i)^{-1} \delta_D(\r_{ij})$.
This survey correlation is ill-defined in Fourier space,
because the selection function $\Phi(\r_i)$ vanishes outside the survey,
so that the Poisson sampling term $\Phi(\r_i)^{-1}$ blows up.
It is only the inverse of the survey correlation,
which leads to the Fisher matrix, which is well-behaved in Fourier space.

\begin{figure}
\vbox to84mm{\rule{0pt}{84mm}}
\includegraphics{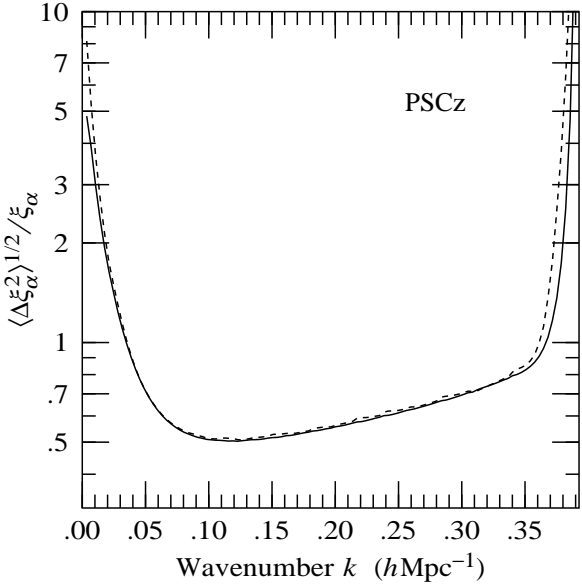}
  \caption[1]{
Expected relative uncertainties
$\langle \Delta\hat\xi_\alpha^2 \rangle^{1/2} / \xi_\alpha$
in the power spectra $\hat\xi_\alpha$ windowed through the kernels
$\psi_\alpha(k)$
of Figure~\protect\ref{goodfish},
plotted against $k_\alpha = \alpha \Delta k$.
The solid line shows the expected uncertainty in the minimum variance estimator,
while the dashed line shows the expected uncertainty in the zeroth order
approximation to the minimum variance estimator,
as described in Paper~1, equations~(97) and (100).
\label{goodvar}
}
\end{figure}

The crowding of the kernels to the right of Figure~\ref{goodfish}
is an artefact of truncating the Fisher matrix at a half-wavelength of
$\PI/(N\Delta k) = 8 \, h^{-1}\Mpc$,
and could easily be avoided by extending the matrix to shorter wavelengths.
I show the crowding to illustrate what happens.
The kernels remain compact, positive, and well-behaved,
with no negative eigenvalues,
if the matrix is extended by a factor of 4 to $N=448$ at the same resolution,
corresponding to a minimum half-wavelength of
$\PI/(N\Delta k) = 2 \, h^{-1}\Mpc$.
Presumably the kernels continue to remain well-behaved beyond this.

The kernels in Figure~\ref{goodfish}
are narrowest and highest at the peak $\sim 0.05 \, h\Mpc^{-1}$
in the prior power spectrum.
The width, hence height, of a kernel in this basis
is approximately equal to the width of the pair density transform $\tildeR(k)$
at the corresponding wavenumber $k$,
which in turn is related to the prior power spectrum as illustrated
in Figure~\ref{wink}.
Increasing the prior power spectrum $\xi(k)$ would narrow the kernel,
while reducing the prior power spectrum would broaden it.
Figure~\ref{wink} shows that, for the PSCz survey,
changing the prior power spectrum by a factor of ten
would change the width of the kernel by a factor of somewhat less than two.
If I had chosen to use as a prior Peacock's (1997) power spectrum
for $\Omega = 0.3$, which is somewhat higher than the adopted case $\Omega = 1$,
then the kernels in Figure~\ref{goodfish} would have been somewhat narrower.

The expected relative uncertainty in the windowed power spectrum
$\hat\xi_\alpha$
is given by
$\langle \Delta\hat\xi_\alpha^2 \rangle^{1/2} / \xi_\alpha$.
Figure~\ref{goodvar}
shows this relative uncertainty for the kernels shown in Figure~\ref{goodfish}.
In addition to the uncertainty in the minimum variance estimator,
Figure~\ref{goodvar} also shows the expected uncertainty
in the zeroth order approximation to the minimum variance estimator,
equations~(97) and (100) of Paper~1.
The two uncertainties are quite similar,
except for the first few kernels at the smallest wavenumbers $k$,
suggesting that the zeroth order approximate estimator
should be adequate except at the longest wavelengths.
The sharp rise in the uncertainty to the right of the graph
(and the discrepancy between the two uncertainties there),
is again an
avoidable artefact of truncating the Fisher matrix at a half-wavelength of
$\PI/(N\Delta k) = 8 \, h^{-1}\Mpc$.
Roughly speaking,
the relative variance of a mode $\alpha$ at wavenumber $\sim k_\alpha$
is given by
$\langle \Delta\hat\xi_\alpha^2 \rangle / \xi_\alpha^2 \approx 2 / (V V_k)$,
where $V$ is the effective volume of the survey at $k_\alpha$,
the volume over which $\xi(k_\alpha) \Phi(\r) \ga 1$,
and $V_k \approx 4\PI k_\alpha^2 \Delta k_\alpha / (2\PI)^3$
is the coherence volume of the mode,
$\Delta k_\alpha$ being the width of the mode.
Interestingly,
the selection function of the PSCz survey falls off in such a way
that the effective volume $V$ initially increases
as the wavelength increases ($k$ decreases)
somewhat more rapidly than the coherence volume $V_k$ decreases,
so causing the relative uncertainty shown in Figure~\ref{goodvar}
to decline mildly with decreasing $k$ down to $\sim 0.1 \, h\Mpc^{-1}$.

\section{Questions}
\label{questions}

This paper leaves unanswered a number of questions
which it would be interesting to investigate further.

\begin{enumerate}
\item
Notwithstanding the mathematical argument at the beginning of \S\ref{compact}
which motivated the idea of applying an infinitely steep scaling function
to the Fisher matrix,
it remains somewhat mysterious as to why the kernels which result
are indeed all positive.
Is positivity true for arbitrary survey geometries?
Is there some fundamental theorem which ensures positivity?
\item
The kernel set obtained from the infinitely steep scaled Fisher matrix
ceased to be positive when the resolution in Fourier space was set too high,
and this breakdown was signalled by the appearance of negative eigenvalues
in the discretized Fisher matrix.
Presumably the appearance of negative eigenvalues is an artefact of
the approximation to the Fisher matrix used here,
since the eigenvalues of the true Fisher matrix should all be
strictly positive.
Nevertheless, the numerical results based on the approximate Fisher matrix
do make physical sense:
somehow the numerics know that the inverse scale of the survey
sets a limit to the resolution at which the power spectrum can be probed.
It would be nice to have a clearer understanding of what is going on here.
\item
Are there other ways to construct sets of kernels which are strictly positive
and yield statistically orthogonal sets of power spectra,
besides applying infinitely steep scaling functions?
Clearly the kernel set illustrated in Figure~\ref{goodfish} is not unique,
since there are other ways of ordering the elements of the scaling function
than the ordering
$\gamma(k_1) \gg \gamma(k_2) \gg \ldots \gg \gamma(k_{N-1})$
adopted there.
I have checked for example that the inverse ordering
$\gamma(k_1) \ll \gamma(k_2) \ll \ldots \ll \gamma(k_{N-1})$
also leads to a strictly positive set of kernels for the PSCz survey.
Whereas the former ordering leads to kernels which have tails to large $k$,
the opposite ordering leads to kernels with tails to small $k$.
\end{enumerate}

I conclude with a caveat about applying the results of this paper
and Paper~1.
The Fisher matrix computed here is for the unredshifted power spectrum,
whereas the galaxy positions in 3-dimensional surveys such as the PSCz survey
lie in redshift space, and are subject to distortions from peculiar velocities.
Although, as explained in \S4 of Paper~1,
the Fisher matrix of the unredshifted power spectrum is the same
irrespective of whether the data lie in real or redshift space,
the minimum variance pair window is not the same in both real
and redshift space.
While there is no harm in applying a pair window which is not minimum
variance (see Paper~1, \S7 for further clarification of this point),
it would be incorrect to carry over the pair window which is minimum
variance in real space and imagine that it is also minimum variance for
the power spectrum in redshift space.
I hope to address the issue of the optimal measurement of the power
spectrum from data in redshift space in a subsequent paper.
Such a measurement would include
a simultaneous measurement of the linear redshift distortion parameter
$\ff \approx \Omega^{0.6} / b$.

\section{Summary}
\label{summary}

This is the second of a pair of papers which
address the problem of how to estimate the unredshifted
power spectrum of fluctuations from a galaxy survey in optimal fashion.

It has been shown that there are many distinct bases of kernels $\psi_\alpha(k)$
which yield statistically orthogonal sets $\hat\xi_\alpha$
of estimates of windowed power spectra.
With each distinct basis is associated a positive definite scaling matrix
$\gamma$.
A basis of kernels is constructed by finding the eigenfunctions
$\phi_\alpha \propto \gamma \psi_\alpha$
of the scaled Fisher matrix $\gamma T \gamma$.

Illustrative results have been presented for the case of the PSCz survey
(Saunders et al., in preparation).
The Fisher matrix of the unredshifted power spectrum of this survey
was computed assuming Gaussian fluctuations and the zeroth order FKP
approximation proposed in Paper~1,
together with the prior power spectrum proposed by Peacock (1997).
Examples of bases of kernels yielding statistically orthogonal sets of
windowed power spectra for this survey have been shown.

The main result of the paper is that,
among the many possible bases of kernels which give rise to statistically
orthogonal sets of windowed power spectra,
there is a particular basis,
obtained by applying an infinitely steep scaling function to the Fisher matrix,
which leads to kernels which are
compact and everywhere positive in Fourier space.
The basis is illustrated for the PSCz survey in Figure~\ref{goodfish}.
This basis is easily computed from the Fisher matrix,
without even the necessity to diagonalize.
The basis of kernels,
along with the associated minimum variance pair weighting described in Paper~1,
\S\S2.3 and 7,
would appear to offer a solution to the problem of how to measure the
unredshifted power spectrum of a galaxy survey in optimal fashion.

\section*{Acknowledgements}

This work was supported by
NSF grant AST93-19977
and by
NASA Astrophysical Theory Grant NAG 5-2797.
I thank Max Tegmark and the referee, Alan Heavens,
for helpful comments.

\end{document}